\documentclass[10pt, letterpaper, twoside, twocolumn, amsmath, amssymb, aps, pra, 
               superscriptaddress, nofootinbib]{revtex4-2}
\usepackage{amsmath}
\usepackage{graphicx}
\usepackage[unicode=true, bookmarks=true, bookmarksnumbered=false,
            bookmarksopen=true, bookmarksopenlevel=1,
            breaklinks=false, pdfborder={0 0 1}, backref=false,
            colorlinks=true]{hyperref}
\hypersetup{pdfborderstyle=, linkcolor=blue, urlcolor=blue, citecolor=blue}
\makeatletter

\usepackage{bm, braket, soul, orcidlink}
\usepackage[normalem]{ulem}

\setstcolor{red}

\makeatother

\begin{document}
\title{Echoes in a parametrically perturbed Kerr-nonlinear oscillator}

\author{Yun-Wen Mao}
\email{ymaoai@student.ubc.ca}
\affiliation{University of British Columbia, Vancouver, B.C. V6T 1Z1, Canada}
\affiliation{Stewart Blusson Quantum Matter Institute, Vancouver, B.C. V6T 1Z4,
Canada.}

\author{Ilia Tutunnikov}
\email{iliatutun@gmail.com}
\affiliation{AMOS and Department of Chemical and Biological Physics, The Weizmann
Institute of Science, Rehovot 7610001, Israel}

\author{Roman V. Krems}
\email{rkrems@chem.ubc.ca}
\affiliation{University of British Columbia, Vancouver, B.C. V6T 1Z1, Canada}
\affiliation{Stewart Blusson Quantum Matter Institute, Vancouver, B.C. V6T 1Z4,
Canada.}

\author{Ilya Sh. Averbukh}
\email{ilya.averbukh@weizmann.ac.il}
\affiliation{University of British Columbia, Vancouver, B.C. V6T 1Z1, Canada}
\affiliation{AMOS and Department of Chemical and Biological Physics, The Weizmann
Institute of Science, Rehovot 7610001, Israel}
\begin{abstract}
We study classical and quantum echoes in a Kerr oscillator driven by a frequency-controlling pulsed perturbation. We consider dynamical response to the perturbation for a single coherent state and for Schr\"{o}dinger cat states constructed as both balanced and imbalanced superpositions of two coherent states. For individual coherent states, we demonstrate that a weak parametric drive yields a long-lived sequence of classical echoes. Cat states are found to exhibit distinct quantum echoes that are sensitive to the initial relative phase and weights of the coherent states in superposition. We examine the effect of dissipation on quantum echoes and quantum revivals of cat states. 
We demonstrate that, even when dissipation suppresses quantum revivals, quantum echoes can be recovered by properly tuning the timing and strength of the perturbation. These results may be useful for characterizing and mitigating errors of cat qubits.
\end{abstract}
\maketitle
Echo appears when a sequence of short pulses perturbs the evolution of an
inhomogeneously broadened ensemble and reverses the dephasing 
dynamics. Echo has been observed in the dynamics of spins~\citep{Hahn_echo_1950,Hahn1953},
cold atoms in optical traps~\cite{Bulatov1998opt, Buchkremer2000opt, Herrera2012opt},
particles in accelerators
~\citep{stupakov1992echo,Spentzouris1996echo,stupakov2013handbook,Sen2018echo},
laser-kicked molecules
~\citep{Karras2015rot_mol_echo,Lin2016rot_mol_echo,Lin2020rot_mol_echo},
optically-resonant media~\cite{Kurnit1964photon_echo,
mukamel1995principles},
plasmas at cyclotron resonance~\cite{Hill1965cyclotron},
electron plasma waves~\cite{Gould1967plasma},
neutron spins~\cite{mezei1972neutron}, 
and free-electron lasers \cite{hemsing2014beam}.
Another type of echo, independent of inhomogeneous broadening, can be realized in single
nonlinear quantum oscillators. Such echoes have been studied in atoms interacting with
a quantized mode of electromagnetic radiation~\citep{Morigi2002,Meunier2005},
a vibrationally excited single molecule~\citep{qiang2020vib_mol_echo}, 
and a quantum Kerr oscillator~\citep{tutunnikov2021echoes}.
Recent experiments have also demonstrated echoes in macroscopically distinguishable
quantum systems, including echoes in a Cooper-pair box~\citep{PhysRevLett.88.047901},
superconducting qubits~\citep{wei2020verifying, PhysRevLett.123.050401, 
kam2024characterization, PhysRevApplied.20.064027},
and trapped-ion qubits~\citep{PhysRevLett.119.150503}.
These experimental results highlight the relevance of echoes in quantum information
processing. In quantum computing applications, echo sequences are used in dynamical
decoupling to prolong qubit coherence~\citep{PhysRevApplied.20.064027, PhysRevLett.119.150503}.
In particular, dynamical decoupling enhances the coherence times
of Greenberger-Horne-Zeilinger (GHZ) and GHZ-like states, which can
be viewed as highly entangled cat states
~\citep{kam2024characterization, PhysRevLett.119.150503}. 

In this paper, we consider a delayed, sudden frequency perturbation to induce echoes in
Kerr oscillators, which differs from previous studies. We identify quantum echoes by
comparing the quantum dynamics to the corresponding classical dynamics. We begin by 
initializing the Kerr oscillator in a single coherent state and show that the timing
and amplitude of the echoes are tunable through the perturbation parameters. We then extend
the study to cat states, and find that the echoes are sensitive to superposition parameters,
including relative weights and phase of the cat components. We show that when dissipation
suppresses quantum revivals, echoes can still be generated by tuning the perturbation timing
and strength to recover observable signatures of coherence. Recent experiments with
superconducting resonators and tunable Kerr nonlinearity have generated highly excited
cat states, offering a platform to probe echoes in cat state dynamics
~\citep{grimm2020stabilization,he2023fast,gravina2023critical}.


The dimensionless Hamiltonian of a quantum Kerr oscillator subject to a sudden frequency
perturbation from a pulsed external field is given by 
(see Appendix~\ref{sec:App-constructing-hamiltonian})
\begin{equation} \label{eq:dimensionless-quantum-H}
\hat{\mathcal{H}} = \hat{n}+\frac{1}{2}\hat{I}+\chi\hat{n}(\hat{n}-\hat{I})
                  - \frac{g(t)}{2}(\hat{a}^{\dagger}+\hat{a})^{2},
\end{equation}
where $\hat{a}^{\dagger}$ and $\hat{a}$ are the creation and annihilation operators, 
$\hat{n}$ is the number operator, $\hat{I}$ is the identity operator, and $\chi$ is the 
dimensionless anharmonicity parameter. The time-dependent interaction term $g(t)$ 
introduces an impulsive frequency change and is defined as $g(t)=g_0\delta(t-\tau)$, 
where $g_0$ is the `kick strength', and the kick is applied at $t=\tau$. Energy and time are
measured in units of $\hbar\omega$ and $1/\omega$, respectively, where $\omega$ is the
fundamental frequency of the oscillator.
For classical dynamics, $\hat{a}$ and $\hat{a}^{\dagger}$ in 
Eq.~\eqref{eq:dimensionless-quantum-H} are replaced by $a=(q+ip)/\sqrt{2}$ and $a^*$, 
respectively, with $q=q/\sqrt{\hbar/(\omega m)}$ and $p=p/\sqrt{\hbar\omega m}$, 
where $q$ is the position, $p$ is the momentum and $m$ is the mass of the oscillator.\\
\begin{figure}[h]
\centering \includegraphics[width = 0.48\textwidth]{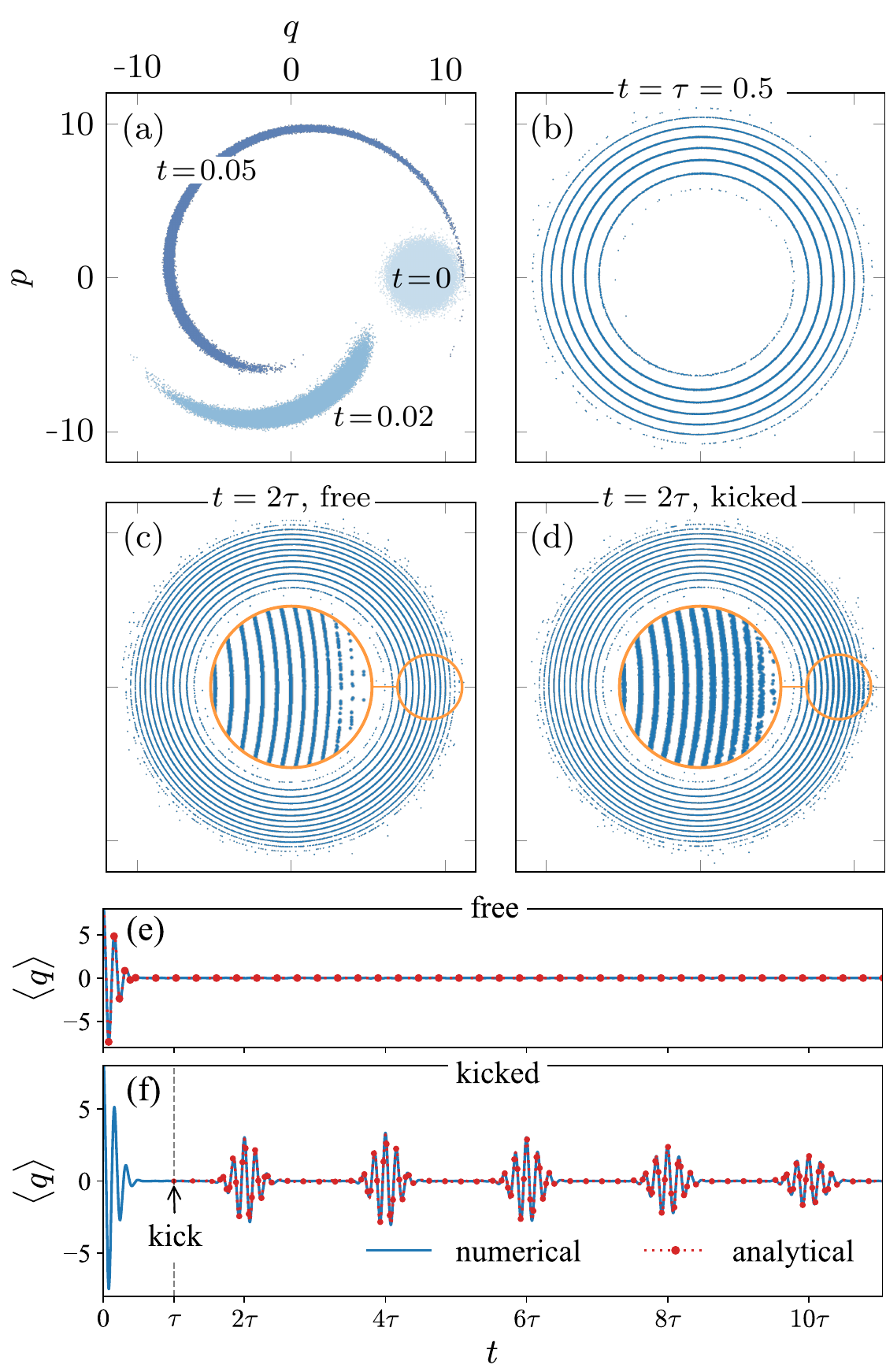} 
\vspace{-1.5em}
\caption{
    (a-d) Phase-space snapshots showing the evolution of $2\times10^{5}$
    classical oscillators. 
    (a) Early stages of filamentation where the initial
    distribution is a Gaussian centered at $(q_0,0)$, corresponding
    to a coherent state $\ket{\alpha_0}$, with $\alpha_0=q_0/\sqrt{2} = 6$. (b)
    Well-developed filaments at $t=\tau=0.5$. 
    (c) Phase-space distribution of the \textit{free} oscillators at $t = 2\tau$. 
    (d) Distribution at the echo time ($t = 2\tau$) following a kick at $t = \tau$.
    The magnified image in (d) highlights thickening filament segments around $q = q_0$
    in contrast to (c).
    (e,f) Average position as a function of time. (e) Free oscillators. 
    (f) Kicked oscillators exhibit echoes at $t = 2n\tau$ 
    ($n\in\mathbb{Z}^+$). Values of the parameters in Eq.~\eqref{eq:dimensionless-quantum-H}
    are $\chi=1$, $g_0=0.01$. The pulse width in numerical simulations is $\sigma_g=10^{-3}$.
    }
\label{fig1} 
\end{figure}

\textbf{\emph{Classical dynamics.}} 
The phase space trajectory of a classical oscillator with the initial position and momentum
$\left[q(0), p(0)\right]$ traces a circle of radius $r=\sqrt{[q(0)]^2+[p(0)]^2}$. 
Thus, $[q(t),p(t)]=R(\Omega t)[q(0),p(0)]$ at time $t$, where $\Omega = 1 + \chi r^2$, and
$R(\cdot)$ is the \textit{clockwise} rotation matrix. Due to circular symmetry, 
$[q,p]=[r\cos(\phi),-r\sin(\phi)]$, where the polar angle $\phi$ increases from the positive
$q$ axis to the negative $p$ axis. In classical dynamics, a coherent state corresponds to 
a Gaussian distribution in phase space, as shown in Fig.~\ref{fig1}(a). The distribution
is centered at $(q, p) = (q_0, p_0)$ with a standard deviation $\sigma=1/\sqrt{2}$, where
$q_0$ and $p_0$ are real numbers. For $p_0 = 0$ and $q_0 > 0$, the time-dependent phase space
distribution is given by
\begin{equation}
    P(\phi,r,t)
    \propto e^{-\frac{q_0^2 + r^2}{2\sigma^2}} 
    \exp\left[\frac{q_0r}{\sigma^2}\cos(\phi-\Omega t)\right].
    \label{eq:P(t)}
\end{equation}
The initial distribution undergoes shearing due to the radius-dependent angular frequency
$\Omega$, with points at larger $r$ rotating faster than those near the origin. Thus,
the initially smooth Gaussian distribution is transformed into a long folded filament
as illustrated in Figs.~\ref{fig1}(a,b). Filamentation is essential for echo formation in
particles accelerators
~\citep{stupakov1992echo, Spentzouris1996echo, stupakov2013handbook, Sen2018echo,
Stupakov_Kauffmann1995echo},
free electron lasers~\citep{hemsing2014beam}, rotating
~\citep{Lin2016rot_mol_echo, Lin2020rot_mol_echo, Karras2015rot_mol_echo, karras2016experimental,
lu2016rot_mol_echo, rosenberg2017rot_mol_echo, rosenberg2018rot_mol_echo}
and vibrating~\citep{qiang2020vib_mol_echo} molecules.

For a large initial displacement $q_0/\sigma \gg 1$, the phase-space distribution evolves
into a tightly wound spiral, as depicted in Fig.~\ref{fig1}(b). The filamented distribution
can be approximated as
\begin{equation}
    P(\phi,r,t)  \propto  e^{-\frac{(r - q_0)^2}{2\sigma^2}}
    \sum_{k=0}^{\infty}e^{-\frac{[r - r_k(\phi)]^2}{2\sigma_k^2(\phi)}},
    \label{eq:P-Gaussian-approx}
\end{equation}
which shows a ring-like distribution centered at $r = q_0$, with a standard deviation
proportional to $\sigma$.  The derivation and the explicit expressions for $r_k(\phi)$ and
$\sigma_k(\phi)$ are given in Appendix~\ref{sec:App-classical-dynamics}.  Each $k$-th term
in Eq.~\eqref{eq:P-Gaussian-approx} corresponds to a revolution of the spiral segment
centered at $r_k(\phi)$, where $\phi \approx \Omega(r)t$. The spacing, $r_{k+1} - r_k$,
decreases over time as $t^{-1/2}$, while the standard deviation $\sigma_k$ narrows as 
$t^{-1/4}$, preserving phase-space volume. As the spiral tightens, $\braket{q(t)}$ decays
according to
\begin{equation} \label{eq:classical-<q>-before-kick}
\braket{q(t)}  = q_0e^{-2q_0^2\sigma^2\chi^2t^2} \cos[(1 + \chi q_0^2)t],
\end{equation}
as illustrated in Fig.~\ref{fig1}(e). Eq.~\eqref{eq:classical-<q>-before-kick} shows that
a larger $q_0$ leads to faster decay.

For kicked oscillators, a weak sudden parametric perturbation ($g_0\ll1$) at $t=\tau$ 
produces echoes in $\braket{q(t)}$ at $t=2n\tau$ ($n\in\mathbb{Z}^+$), as shown in 
Fig.~\ref{fig1}(f). This perturbation preserves oscillator positions but updates their momenta
$p(\tau_+) = p(\tau_-) + 2g_0q(\tau_-)$, where $\tau_-$ and $\tau_+$ denote the times 
immediately before and after the kick, respectively.
For weak kicks, the instantaneous phase-space displacement is negligible, but the effect
of angular frequency shift, 
$\Omega_{+}=\Omega+4\chi g_0(qp + gq)\approx\Omega-2\chi g_0\sin(2\phi)r^2$, becomes 
significant over time.
The kick reduces the angular frequency when $\sin2\phi>0$, and increases it when $\sin2\phi<0$. 
This kick-induced correction has the same $r^2$-dependence
as the original Kerr-induced frequency shift responsible for the phase-space shearing.
After an additional time $\tau$, oscillators in the spiral filament \textit{partially reverse}
the phase shift accumulated before the kick. Thus, at $t=2\tau$, many oscillators realign
near $q=q_0$, locally thickening the otherwise thinned spiral which manifests in the first
macroscopic \emph{echo} [compare Figs.~\ref{fig1}(c,d)]. This partial re-synchronisation
of the oscillators recurs at multiples of $\tau$. Depending on the phase-space symmetry of
the observable, echoes may appear at integer multiples or fractions of $\tau$.

The average position after the kick is given by
\begin{equation} \label{eq:classical-<q>-after-kick}
\braket{q_+(t>\tau)} =\sum_{n=1}^\infty \int\limits_0^\infty h(r,t)e^{i\Omega(t-2n\tau)}\,dr,
\end{equation}
where $h(r,t)$ is a slowly varying function of time localized around $r=q_0$. 
Due to the rapidly oscillating exponent, the integral gets significant contributions only
when the phase is stationary, i.e., at $t\approx2n\tau$,
as demonstrated in Fig.~\ref{fig1}(f). The first echo is approximated as
\begin{equation}\label{eq:classical-<q>-at-2tau}
    \braket{q_+(t\approx2\tau)}\approx J_1[2\chi\tau g_0 q_0^2]\braket{q(t-2\tau)},
\end{equation}
where $J_1$ is a Bessel function of the first kind. 
Notably, the echo amplitude 
depends on $\tau g_0q_0^2$, while the timing is independent of both $g_0$ and $q_0$.\\

\begin{figure}[!tp]
\centering \includegraphics[width = 0.48\textwidth]{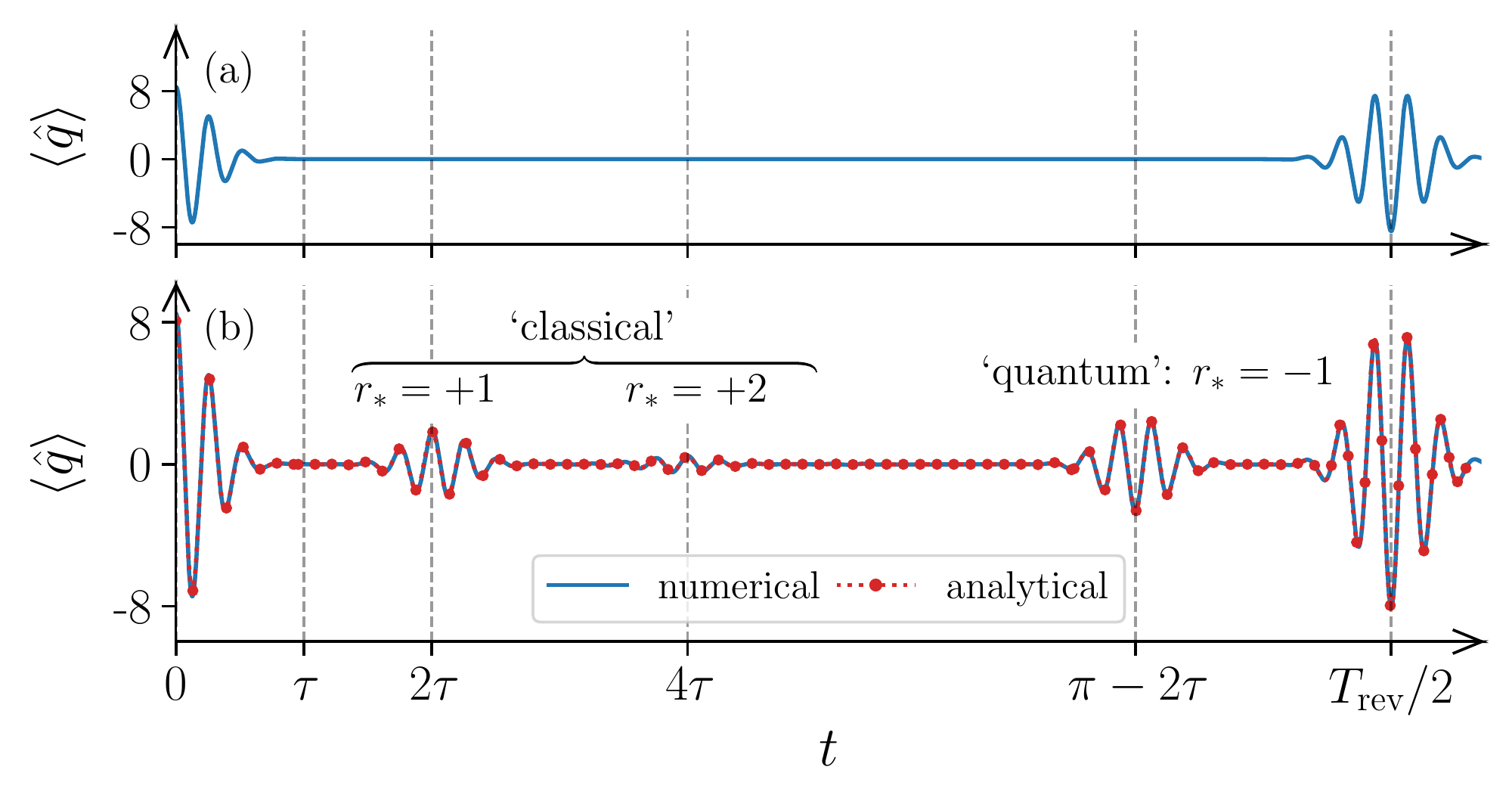} 
\vspace{-2.5em}
\caption{Quantum position expectation value as a function of time for the (a) free and (b)
    kicked Kerr oscillator. The initial state is a coherent state, $\ket{\alpha_0}$ with 
    $\alpha_0 = 6$. The values of the parameters for Eqs.~\eqref{eq:dimensionless-quantum-H}
    and~\eqref{eq:quantum-<q>-kicked} are $\chi=1$, $\tau=2\pi/19\approx0.33$, and 
    $g_0=0.01$. The pulse width in numerical simulations is $\sigma_g=10^{-3}$.}
\label{fig2}
\end{figure}

\textbf{\emph{Quantum dynamics.}} 
The discrete spectrum gives rise to revivals and a distinct quantum echo sequence.
We begin by analyzing the quantum dynamics of a coherent initial state $\ket{\alpha_0}$,
where the complex number $\alpha_0$ defines both the amplitude and phase of the state. 
The expectation value $\braket{\hat{q}(t)}$ for a free oscillator is given by 
$\braket{\hat{q}(t)}=\sqrt{2}\mathrm{Re}[\braket{\alpha_0|\hat{a}(t)|\alpha_0}]$,
where
\begin{equation} \label{eq:quantum-<q>-free}
\braket{\hat{a}(t)} = \alpha_0 e^{-|\alpha_0|^2}e^{-it}
                      \exp\left(|\alpha_0|^2e^{-i2\chi t}\right).
\end{equation}
At short times, $\braket{\hat{q}(t)}$ agrees with the classical result in 
Eq.~\eqref{eq:classical-<q>-before-kick}, with $q_0=\sqrt2\alpha_0$ and $\sigma=1/\sqrt{2}$, 
showing decaying oscillations (Fig.~\ref{fig2}). 
At longer times, the quadratic energy spectrum leads to quantum revivals.
For the Kerr oscillator, with energy $E_{n}=n+1/2+\chi n(n-1)$,  where $n\in\mathbb{Z}^+$,
the revival time is $T_\text{rev}=4\pi(d^{2}E_n/dn^{2})^{-1}=2\pi/\chi$ 
(for a review see, e.g.~\citep{Robinett2004}).
Observables may also exhibit \emph{fractional revivals} at 
$t=T_\text{rev}/\nu$, with $\nu>1\in\mathbb{Z}^{+}$ \cite{AVERBUKH1989449}. 
For example, Fig.~\ref{fig2}(a) shows a half revival of $\braket{\hat{q}(t)}$ at
$t=T_\text{rev}/2=\pi$ \cite{kirchmair2013kerr_revival, Herrera2012revival}.

An impulsive perturbation induces an echo at $t=2\tau$ [see Fig.~\ref{fig2}(b)], similar
to the echo in classical dynamics. However, the discrete energy spectrum of the quantum
oscillator also leads to a second sequence of quantum echoes appearing \emph{before} the
first revival. For $\braket{\hat{q}(t)}$, these echoes occur at $t=T_{\text{rev}}/2-2\tau n$.
Unlike the fixed timing of quantum revivals, the timing of these echoes is tunable.
The echoes are described by
\begin{equation} \label{eq:quantum-<q>-kicked}
\begin{aligned}
\braket{\hat{a}(t>\tau)}_{r_*} & \approx J_{r_*}[iz(t)]\braket{\hat{a}(t - 2r_*\tau)},\\
z(t) & =2\alpha_0^2g_0\left[e^{-i2\chi(t-\tau)} - \cos(\chi l\tau)\right],
\end{aligned}
\end{equation}
where $l$ is a non-negative integer, $\braket{\hat{a}}$ is given by 
Eq.~\eqref{eq:quantum-<q>-free}, and $J_{r_*}$ is the Bessel function of order $r_*$. 
This expression is valid for $\tau=T_\text{rev}/\nu$ with $\nu$ odd and 
$\nu\,\text{mod}\,4=3$. When $\tau$ is small compared to $T_\text{rev}$, the index 
$r_*$ selects `classical' $(r_* > 0)$ and `quantum' $(r_* < 0)$ echoes of order $r_*$. 
The relation, $l=(4r_*-2)\text{ mod }\nu$, fixes the value of $l$ corresponding to each 
$r_*$. Echo amplitudes are given by $|J_{r_*}[iz(2r_*\tau)]|$. For small $\tau$, the 
amplitude of the first echo ($r_*=1,\,m=2$) is approximated by 
$J_1[g_0 q_0^2\sin(2\chi\tau)]\approx J_1(2\chi\tau g_0 q_0^2)$,
as in the classical case in Eq.~\eqref{eq:classical-<q>-at-2tau}. Note that the echoes
depend only on the total area of the pulse, not on the pulse shape, provided the 
echo-inducing pulse is short.\\

\emph{Two-component cat states.} We now consider the dynamics of a two-component cat state defined as
\begin{equation} \label{eq:cat-state-definition}
\ket{\mathcal{C}}=\mathcal{N}_+ \ket{+\alpha_0} + \mathcal{N}_- e^{i\theta}\ket{-\alpha_0},
\end{equation}
where $\theta$ is the relative phase, and $\mathcal{N}_{\pm}>0$ are the probability amplitudes acting on the coherent states $\ket{\pm\alpha_0}$.
States with $\mathcal{N}_+ = \mathcal{N}_-$ are labeled $\ket{\mathcal{C}_{\rm sym}}$,
while states with $\mathcal{N}_+ \neq \mathcal{N}_-$ are labeled $\ket{\mathcal{C}_{\rm asym}}$. 
The expectation value $\braket{\hat{q}}$ for the freely evolving cat state is determined by
\begin{align} \label{eq:cat-<q(t)>-explicit}
\braket{\mathcal{C}|\hat{a}|\mathcal{C}} 
& =\alpha_0e^{-it}e^{-|\alpha_0|^2}
   \left\{(\mathcal{N}_+^2 - \mathcal{N}_-^2)\exp\left[|\alpha_0|^2 e^{-i2\chi t}\right]\right. \nonumber \\
 & \left.-i2\mathcal{N}_+\mathcal{N}_-\sin(\theta)\exp\left[-|\alpha_0|^2 e^{-i2\chi t}\right]\right\}.
\end{align}
The first term has the same time dependence as in Eq.~\eqref{eq:quantum-<q>-free} and
produces half revivals at $t=nT_{\text{rev}}/2=n\pi$. The second term has the opposite sign
in the exponent and produces quarter revivals at $t=nT_\text{rev}/4=n\pi/2$, which are
unique to cat states. Note that the amplitude of the quarter revival is proportional to 
$\sin(\theta)$.
\begin{figure}[!tp]
\centering 
\includegraphics[width=0.48\textwidth] {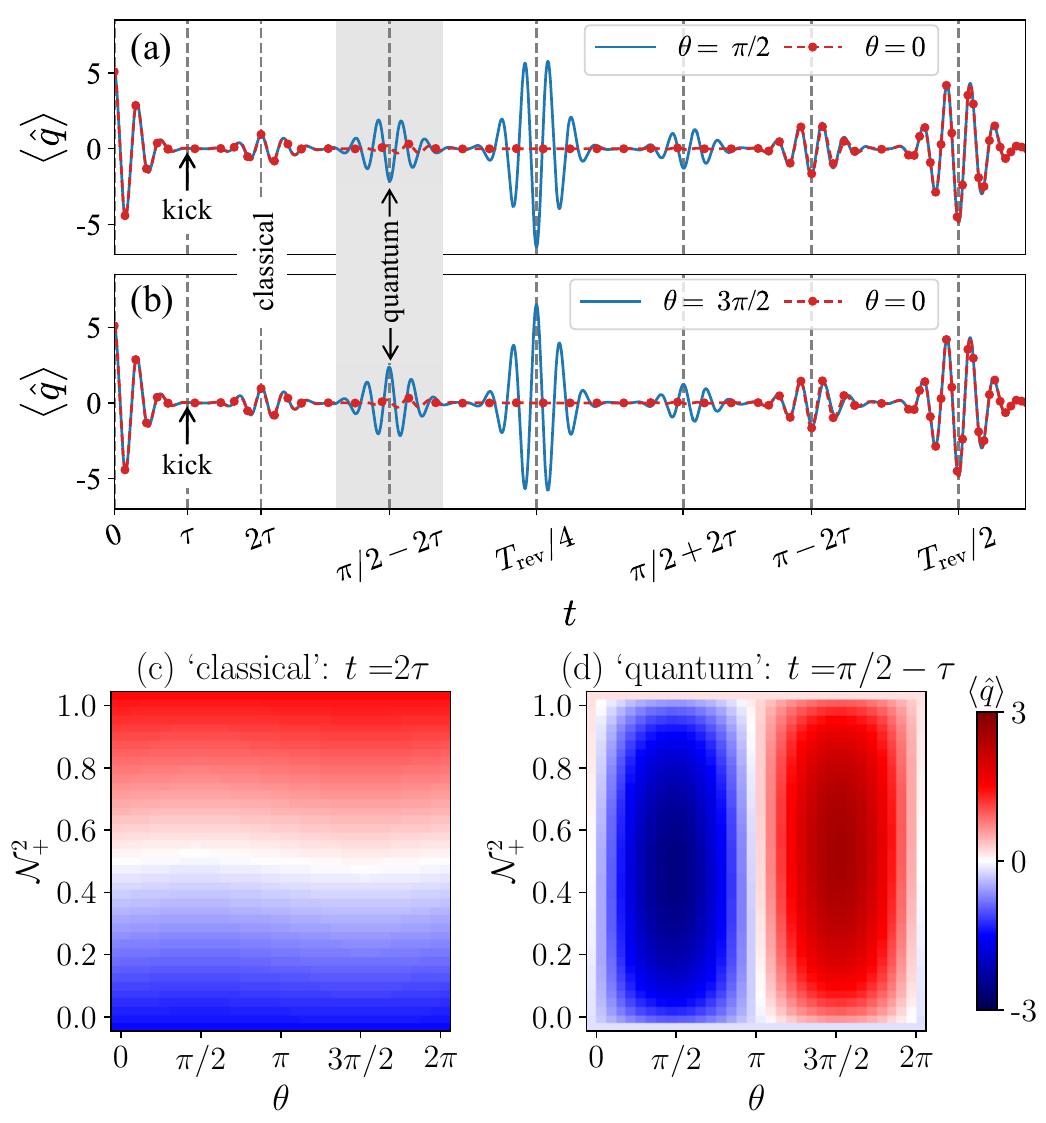} 
\vspace{-2em}
\caption{(a,b) Position expectation values of kicked asymmetric 
    cat states ($\mathcal{N}_+^2 = 1 - \mathcal{N}_-^2 = 0.8$, $\alpha_0=6$). 
    (c,d) Amplitudes of the first classical and quantum echoes as a function of the initial
    cat state parameters, $\mathcal{N}_+$ and $\theta$. Other fixed parameters in 
    Eq.~\eqref{eq:dimensionless-quantum-H}: $\chi=1$, $\tau=0.27$, $g_0=0.01$.
    }
\label{fig3}
\end{figure}
Figs.~\ref{fig3}(a,b) show examples of $\braket{\hat{q}}$ for kicked asymmetric cats. 
As before, the first classical echo emerges at $t=2\tau$, while
the first pure quantum echo appears \emph{before the quarter revival}, at $t=\pi/2-2\tau$.
As demonstrated by Eq.~\eqref{eq:cat-<q(t)>-explicit}, the sign of the quarter revival
is determined by $\sin(\theta)$, while the sign of the half revival is insensitive to 
$\theta$.

For a more comprehensive picture, we plot the amplitudes of the first classical and quantum
echoes as functions of $\theta$ and $\mathcal{N}_+$ in Fig.~\ref{fig3}(c,d). 
The classical echo amplitude is sensitive to $\mathcal{N}_+$, while the quantum echo
amplitude resembles that of the quarter revival, showing a sinusoidal dependence on $\theta$
and non-monotonic dependence on $\mathcal{N}_+$. The sensitivity of the cat echoes
to the superposition probability amplitudes $\mathcal{N}_{\pm}$ and the relative phase $\theta$ suggests applications for state
characterization. The symmetry of $\ket{C_\text{sym}}$  forbids the classical echo. 
The quantum echo before $T_{\rm rev}/4$ also vanishes with the disappearance of the revival,
which occurs when $\theta=0,\pi,2\pi\implies\sin(\theta)=0 $.\\

\textbf{\emph{Dissipation effects.}} We consider dissipation effects due to interactions
with a reservoir at finite temperature $T$. The reduced density matrix of
the oscillator $\hat{\rho}$ obeys the Lindblad master equation \citep{Milburn1986}
\begin{equation} \label{eq:master-equation}
\dot{\hat{\rho}} = -i[\hat{\mathcal{H}},\hat{\rho}]
+
\gamma\bar{n}[[\hat{a},\hat{\rho}],\hat{a}^{\dagger}]
+
\frac{\gamma}{2}(2\hat{a}\hat{\rho}\hat{a}^{\dagger}- \{\hat{a}^{\dagger}\hat{a},\hat{\rho}\}),
\end{equation}
where $\{\hat{A},\hat{B}\} = \hat{A}\hat{B} + \hat{B}\hat{A}$, $\hat{\mathcal{H}}$ is the
Hamiltonian in Eq.~\eqref{eq:dimensionless-quantum-H}, $\gamma$ is 
the damping constant, and $\bar{n}$ is the mean number of thermal excitations in the reservoir
at a given frequency $\omega$, 
described by the Bose-Einstein distribution $\bar{n}=[\exp(\epsilon)-1]^{-1}$, where 
$\epsilon=\hbar\omega/(k_{B}T)$ and $k_{B}$ is the Boltzmann constant.

\begin{figure}[!tp]
\centering 
\includegraphics[width=0.49\textwidth]{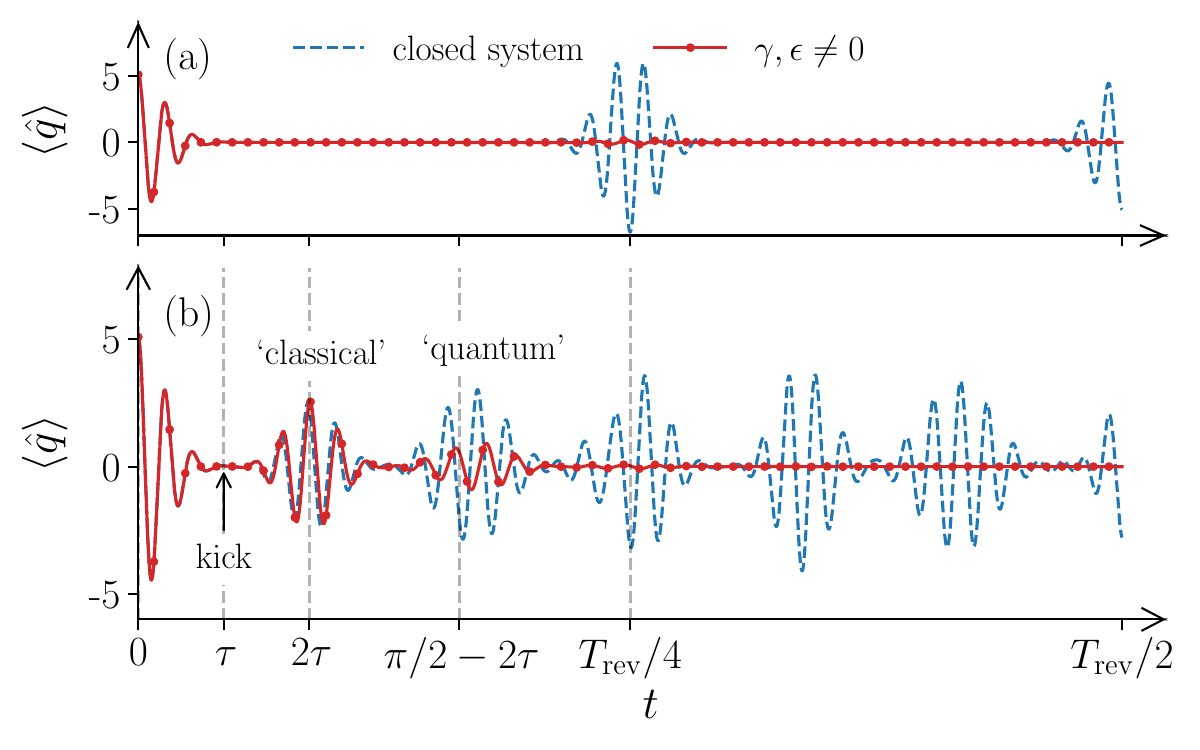}
\vspace{-2em}
\caption{Position expectation value as a function of time for (a) a free and (b) a kicked
asymmetric cat state ($\mathcal{N}_+^2=1-\mathcal{N}_-^2=0.8$, $\alpha_0=6$, $\theta=\pi/2$)
with or without dissipation. Common parameters in Eq.~\eqref{eq:master-equation}: $\chi=1$,
damping constant $\gamma=0.03$, and dimensionless temperature $\epsilon^{-1}=1$. 
Kick parameters in (b): $\tau=0.27$, $g_0=0.03$.}
\label{fig4}
\end{figure}

At $T=0$ (i.e., $\bar{n}=0$), Eq.~\eqref{eq:master-equation} simplifies and admits an
analytical solution for the freely evolving oscillator~\citep{Arevalo-Aguilar2008}.
We find that damping can be accounted for by replacing $2i\chi$ with $2i\chi+\gamma$ in
Eqs.~\eqref{eq:quantum-<q>-free} and \eqref{eq:cat-<q(t)>-explicit}.
Dissipation causes exponential suppression of the revivals by a factor of
$\exp[|\alpha_0|^2(e^{-\gamma t}-1)]\approx\exp[-|\alpha_0|^2\gamma t]$.
Notably, this suppression is significant for cat states with high amplitude components
($|\alpha_0|^2 \gg 1$) even when dissipation is weak ($\gamma \ll 1$). Finite temperature
further accelerates signal decay. For example, Fig.~\ref{fig4}(a) shows that for 
$\alpha_0 = 6$, with $\gamma=0.03$ and $\epsilon^{-1} = 1$, the amplitude of the first
quarter revival is negligible.

Since the effect of dissipation grows exponentially with time, time-tunable echoes can be
used to retrieve coherence signatures before significant decay. Fig.~\ref{fig4}(b) compares
echo and revival dynamics of a kicked cat state with or without dissipation. Notably, 
the first `classical' ($t=2\tau$) and `quantum'  ($t=\pi/2-2\tau$) echoes remain clearly visible
despite dissipation and damping, while the quarter revival is suppressed by dissipation.
Thus, echoes may be a promising tool for probing quantum coherence in cat states under
dissipative conditions.\\

In summary, we analyzed the classical and quantum echoes induced by
a sudden frequency perturbation in a Kerr oscillator.
The dynamics of a single coherent state exhibits a long-lived sequence of classical echoes,
markedly different from that of a linearly excited Kerr
oscillator studied earlier~\citep{tutunnikov2021echoes}. For quantum oscillators,
an additional sequence of quantum echoes emerges prior to the first quantum revival,
and their amplitudes can exceed those of classical echoes.

The dynamics of two-component cat states exhibit a quarter-revival, in contrast to a single
coherent state, which undergoes only a half-revival at a later time. The difference 
in revival dynamics causes echoes in cat states to appear earlier than in individual
coherent states. The timing and amplitude of these echoes are tunable through perturbation
parameters. This tunability suggests potential applications in dynamical decoupling, where
sequences of frequency kicks can be applied to idle qubits to mitigate decoherence. A key
challenge in dynamical decoupling arises from imperfections in the control pulses, as errors
can accumulate over time~\citep{tyryshkin2010dynamicaldecouplingpresencerealistic}.
In this context, our work demonstrates that echoes are robust against pulse-shape
imperfections, provided that the pulses are short and the total pulse area is preserved.
Under dissipation that eliminates quantum revivals, we show that tuning the timing and
strength of the perturbation allows the recovery of quantum echoes. This can be used to
study decoherence or for tomography of cat states. Such tomography can be facilitated 
by machine learning approaches, such as Bayesian optimization, as illustrated in our
previous work for inverse problems based on time-dependent quantum dynamics~\citep{krems2019bayesian}. 

\begin{acknowledgments}
This work was supported by NSERC of Canada. I.A. gratefully acknowledges the hospitality
extended to him during his stay at the Department of Chemistry of the University of
British Columbia. This research was made possible in part by the historic generosity of
the Harold Perlman Family. 
\end{acknowledgments}

\bibliography{bibliography}

\appendix
\section{Derivation of the  Dimensionless Hamiltonian}
\label{sec:App-constructing-hamiltonian}
We start with the Hamiltonian of a freely evolving harmonic oscillator 
\begin{equation}\label{Equation: free HO px}
    \hat{H}_{\rm free, HO} = \frac{\hat{p}^2}{2m} + \frac{1}{2} m\omega^2 \hat{q}^2, 
\end{equation}
where $m$ is the mass, $\omega$ is the oscillator frequency, $\hat{p}$ is the momentum
operator and $\hat{q}$ is the position operator. The Hamiltonian of a kicked harmonic
oscillator is given by
\begin{equation}\label{Equation: squeezed HO px}
    \hat{H}_{\rm kicked, HO} = \frac{\hat{p}^2}{2m} 
                             + \frac{[\omega^2-2\Gamma(t)]}{2} m \hat{q}^2.
\end{equation}
Here the time-dependent interaction term $\Gamma(t)$ is included, which induces a parametric
excitation. Substituting $\hat{p}$ and $\hat{q}$ with the annihilation 
($\hat{a} =\sqrt{m\Omega/2\hbar}\left[\hat{x} + i\hat{p}/(m\omega)\right]$) and creation
($\hat{a}^{\dagger}$) operators allows us to re-write Eq.~\eqref{Equation: squeezed HO px} as
\begin{equation}\label{Equation: free HO aad}
    \hat{H}_{\rm kicked, HO} \!=\! -\frac{\hbar \omega}{4}(\hat{a}^{\dagger} \! -  \hat{a})^2 
                             + \frac{[\omega^2-2\Gamma(t)]\hbar }{4\omega}
                               (\hat{a}^\dagger + \hat{a})^2.
\end{equation}
The Kerr-nonlinearity term  $\hbar K (\hat{a}^{\dagger})^2\hat{a}^2$ is then included to
the Hamiltonian, giving
\begin{align} \label{Equation: derivision of dimensionfull squeezed Kerr H}
\hat{H}_{\rm kicked, Kerr} & =  \hbar\omega(a^{\dagger}a+\frac{1}{2})
                             + \hbar K(\hat{a}^{\dagger})^2\hat{a}^2 \nonumber \\ 
                           & - \frac{\Gamma(t)\hbar}{2\omega}(\hat{a}^\dagger + \hat{a})^2,
\end{align}
where $K$ is the anharmonicity parameter. The dimensionless Hamiltonian in 
Eq.~\eqref{eq:dimensionless-quantum-H} is obtained by introducing the dimensionless 
parameters $\chi = K/\hbar \omega$ and $g(t) = \Gamma(t)/\hbar\omega^2$, and expressing
energy in units of $\hbar\omega$. Substituting the number operator $\hat{n} = a^\dagger a$,
the dimensionless Hamiltonian is given as
\begin{equation}\label{Equation: derivision of dimensionless squeezed Kerr H}
    \hat{\mathcal{H}} = \hat{n}+\frac{1}{2}\hat{I}+\chi\hat{n}(\hat{n}-\hat{I})
                  - \frac{g(t)}{2}(\hat{a}^{\dagger}+\hat{a})^{2},
\end{equation}
where $\hat{I}$ is the identity operator.

\section{Classical Dynamics \label{sec:App-classical-dynamics}}

This section includes the derivations of (i) the exact classical solutions for a single 
Kerr-nonlinear oscillator, (ii) classical time-dependent phase space distribution for 
an ensemble of Kerr-oscillators, (iii) classical ensemble-averaged position for the free
and kicked oscillators [see Eqs.~\eqref{eq:classical-<q>-before-kick}~and
~\eqref{eq:classical-<q>-after-kick}]. Additionally, we briefly describe our numerical
approach for the classical ensemble simulations.\\

\textbf{\emph{Exact classical solutions.}} The classical Hamiltonian in terms of $q$ and
$p$ is given by
\begin{equation}\label{eq:dimensionless-classical-H}
\mathcal{H} = \frac{1}{2}(q^2 + p^2) + \frac{1}{2}+\frac{\chi}{4}(q^2 + p^2)^2
            - g(t)q^2.
\end{equation}
The position and momentum of a single free classical Kerr oscillator satisfy
the Hamilton's equations
\begin{equation}
\begin{aligned}
\dot{q} & = \frac{dH}{dp} = \chi(q^2p + p^3) + p = p(1+\chi r^2),\\
\dot{p} & = -\frac{dH}{dq} = -\chi(q^3 + qp^2) - q = -q(1+\chi r^2),
\end{aligned}
\end{equation}
where $r^{2}=q^{2}+p^{2}$. Note that $r$ is a constant of motion,
since 
\begin{align}
\frac{d(r^2)}{dt} & = \frac{dr}{dq}\frac{dq}{dt}
                      + \frac{dr}{dq}\frac{dp}{dt} \nonumber \\
                    & = 2qp(\chi r + 1) - 2pq(\chi r+1) = 0.
\end{align}
The second derivative of the position is given by
\begin{equation}
\ddot{q} = \frac{d}{dt}[p(\chi r + 1)] = -q(1 + \chi r^2)^2,
\end{equation}
which has the exact solution
\begin{equation}\label{eq:App-classical-q}
q(t)=A\cos(\Omega t+\phi_0),
\end{equation}
where $A$ and $\phi_0$ are determined by the initial conditions,
and $\Omega=1+\chi r^{2}$. Similarly, $p(t)$ is given by
\begin{equation}
p=\frac{\dot{q}}{\Omega}=-A\sin(\Omega t+\phi_{0}).\label{eq:App-classical-p}
\end{equation}
Using Eq.~\eqref{eq:App-classical-q}~and~\eqref{eq:App-classical-p},
we can find $A$ and $\phi_0$ explicitly. At $t=0$, the initial
position and momentum are $q(0)$ and $p(0)$, respectively.

This yields $q(0)=A\cos(\phi_{0})$ and $p(0)=-A\sin(\phi_{0})$,
\begin{equation}
\begin{aligned}
\frac{q(0)}{\cos(\phi_{0})} & =-\frac{p(0)}{\sin(\phi_{0})},\\
\phi_{0} & = \arctan[-p(0)/q(0)],
\end{aligned}
\end{equation}
and
\begin{equation}
q(0)^2 + p(0)^2 = A^2\cos^2{(\phi_{0})} + A^2\sin^2{(\phi_{0})},
\end{equation}
yielding $A=r=\sqrt{q(0)^2 + p(0)^2}$. Finally, the expressions
for $q(t)$ and $p(t)$ are written in terms of the initial conditions
as
\begin{equation}
\begin{aligned}
q(t) & = \sqrt{r}\cos\left\{\Omega t + \arctan[-p(0)/q(0)]\right\},\\
p(t) & = -\sqrt{r}\sin\left\{\Omega t + \arctan[-p(0)/q(0)]\right\}.
\end{aligned}
\end{equation}

Alternatively, $q(t)$ and $p(t)$ can be written as
\begin{equation}
\begin{aligned}
q(t) & = q(0)\cos{(\Omega t)} +p(0)\sin{(\Omega t)},\\
p(t) & = p(0)\cos{(\Omega t)} - q(0)\sin{(\Omega t)}.
\end{aligned}
\end{equation}
or
\begin{equation}
    \begin{bmatrix}
    q(t)\\
    p(t)
    \end{bmatrix}
    =R(\Omega t)
    \begin{bmatrix}
    q(0)\\
    p(0)
    \end{bmatrix}
\end{equation}
where $R(\cdot)$ is the standard rotation matrix that rotates points in the phase space
clockwise.\\

\textbf{\emph{Time-dependent phase space distribution.}}
Classical phase space distribution corresponding to a coherent state 
$\ket{\alpha_0}=\ket{q_0/\sqrt{2}+ip_0/\sqrt{2}}$ is
\begin{equation} \label{eq:App-initial-phase-space}
f\left[q(0),p(0)|\alpha_{0}\right] 
=\frac{1}{2\pi\sigma^2} e^{-\frac{[q(0) - q_0]^{2}}{2\sigma^2}}
e^{-\frac{[p(0) - p_0]^{2}}{2\sigma^2}}
\end{equation}
where $\sigma = 1/\sqrt{2}$ for a non-squeezed coherent state. To obtain the time-dependent
distribution (before the kick), we invert the relation between $[q(0),p(0)]$ and 
$[q(t),p(t)]$, i.e., $[q(0),p(0)]=[R(\Omega t)]^{-1}[q(t),p(t)]=R(-\Omega t)[q(t),p(t)]$%
\footnote{The inverse of a clockwise rotation by an angle $\Omega t$ is an opposite
rotation by the same angle. That is why $[R(\Omega t)]^{-1}=R(-\Omega t)$.} 
and substitute these into Eq.~\eqref{eq:App-initial-phase-space}. Simplification results
in Eq.~\eqref{eq:P(t)}.\\

\textbf{\emph{Gaussian approximation of the phase space density.}}--
The phase space density in Eq.~\eqref{eq:P(t)} can be approximated
by the second exponential term to second order around the zeros of $\phi-\Omega(r)t$, $r_{k}$.
The resulting explicit expression reads:
\begin{equation}
\begin{aligned}P(\phi,r,t) & \approx\frac{e^{-\frac{(r-q_{0})^{2}}{2\sigma^{2}}}}{2\pi\sigma^{2}}\\
 & \times\sum_{k=0}^{\infty}\exp{\left[-\frac{2q_{0}r_{k}^{3}\chi^{2}t^{2}}{\sigma^{2}}(r-r_{k})^{2}\right]},\\
\end{aligned}
\label{eq:App-P-Gaussian-approx-explicit}
\end{equation}
where
\begin{equation}\label{eq:r_k}
    r_{k}(\phi)  =\frac{1}{\sqrt{\chi}}\sqrt{\frac{2\pi k+\phi}{t}-1}.
\end{equation}
Introducing the expression
\begin{equation}
    \sigma_k(\phi) = \frac{\sigma}{2\sqrt{q_0}\chi^{1/4}t^{1/4}(2\pi k + \phi -t)^{3/4}},
\end{equation}
allows us to simplify Eq.~\eqref{eq:App-P-Gaussian-approx-explicit} to
\begin{equation}
    P(\phi,r,t) \cdot 2\pi\sigma^2 \approx  e^{-\frac{(r - q_0)^2}{2\sigma^2}}
    \sum_{k=0}^{\infty}e^{-\frac{[r - r_k(\phi)]^2}{2\sigma_k^2(\phi)}}.
\end{equation}

\textbf{\textit{Average position, free propagation.}} First, we
apply the Jacobi-Anger expansion in Eq.~\eqref{eq:P(t)} 
\begin{align}
\exp\left[\frac{q_0 r}{\sigma^2}\cos(\phi-\Omega t)\right] & =I_0\left[\frac{q_0 r}{\sigma^{2}}\right]\nonumber \\
+2\sum_{m=1}^{\infty}I_{m}\left[\frac{q_0 r}{\sigma^2}\right] & \cos[m(\phi-\Omega t)].
\label{eq:App-P-Jacobi-Anger}
\end{align}
where $I_{0,m}$ are the modified Bessel functions of the first kind.
Then, the $d\phi$ integration reduces $\braket{q(t)}$ to
\begin{align}
\braket{q(t)} & =\frac{1}{\sigma^2}\int_{0}^{\infty}I_1\left[\frac{q_0r}{\sigma^{2}}\right]
e^{-\frac{q_0^2+r^2}{2\sigma^2}}\cos(\Omega t)r^2\,dr\nonumber \\
 & \sim\frac{1}{\sigma\sqrt{2\pi q_0}}\int_0^\infty e^{-\frac{(r-q_0)^{2}}{2\sigma^2}}
   \cos(\Omega t)r^{3/2}\,dr,
\end{align}
where we used the asymptotic form of $I_{1}(x)\sim e^{-x}/\sqrt{2\pi x}$.

The integral can be approximated by replacing $r^{3/2}\rightarrow q_0^{3/2}$,
since this factor only slightly varies in the relevant integration range. Then,
the integral reads
\begin{align}
\braket{q(t)} & \approx\frac{q_0}{\sigma\sqrt{2\pi}}\int_0^\infty 
                e^{-\frac{(r-q_0)^2}{2\sigma^2}}\cos(\Omega t)\,dr\nonumber \\
              & \approx q_0 e^{-2q_0^2\sigma^2 \chi^2 t^2}\cos[(1+\chi q_0^2)t],
\label{eq:App-average-position}
\end{align}
where we also assumed $4\sigma^4 \chi^2 t^2\ll1$ to obtain a more compact form.\\

\textbf{\emph{Numerical simulations.}}
The numerical results in Fig.~\ref{fig1} were obtained by the Monte Carlo approach.
The initial conditions [$q(0)$ and $p(0)$] were sampled from the distribution in 
Eq.~\eqref{eq:App-initial-phase-space}. The trajectories were obtained by numerically
solving Hamilton's equations
\begin{equation} \label{eq:classical-eq-of-motion}
\begin{aligned}
\frac{dq}{dt} & =\frac{dH}{dp} = \chi(q^2 p + p^3)+p,\\
\frac{dp}{dt} & =-\frac{dH}{dq} = -\chi(q^3 + qp^2) - q + 2g(t)q.
\end{aligned}
\end{equation}
For numerical calculations, $g(t)$ in Eqs.~\eqref{eq:dimensionless-quantum-H} 
and~\eqref{eq:dimensionless-classical-H} was approximated by a narrow in time Gaussian function:
\begin{equation} \label{eq:App-Gauss-gt}
g(t)_{\rm approx} = \frac{g_0}{\sqrt{2\pi}\sigma_g}\exp\left[-\frac{(t-\tau)^2}{2\sigma_g^2}\right],
\end{equation}
where $\sigma_{g}\ll1/\omega$.
Specifically, these ODEs were solved with the Python package \texttt{SciPy}
\citep{2020SciPy-NMeth}.\\

\textbf{\textit{Average position after the kick.}}
Assuming $g_0\ll1$, the position of an oscillator after the weak impulsive kick is given by
\begin{equation}
\begin{aligned}
q_{+}(t) 
& \approx 
r\cos(\phi+\Omega_+t)=r\mathrm{Re}[e^{i\phi}e^{i\Omega_+t}]\\
 & =r\mathrm{Re}[e^{i\phi}e^{i\Omega t}e^{-2i\chi g_0\sin(2\phi)r^2 t}]\\
\Omega_+ & \approx\Omega-2\chi g_0\sin(2\phi)r^2.
\end{aligned}
\end{equation}
Next, we use the Jacobi-Anger expansion twice, for the phase space
density, like in Eq.~\eqref{eq:App-P-Jacobi-Anger}, and for 
\begin{equation}
e^{-2i\chi g_0\sin(2\phi)r^{2}t}=\sum_{n=-\infty}^{\infty}J_{n}(2\chi g_0r^{2}t)e^{-i2n\phi}.
\end{equation}
These expansions allow us to carry out the $d\phi$ integration in
Eq.~\eqref{eq:classical-<q>-after-kick} exactly (it is understood
that we need the real part of the following expression) 
\begin{align}
\braket{q_+(t)} 
& =
\frac{1}{\pi\sigma^2}\int_0^{\infty}\sum_{m=1}^{\infty}\sum_{n=-\infty}^{\infty}
I_m\left[\frac{q_0 r}{\sigma^2}\right] J_n(2\chi g_0 r^2 t)e^{i\Omega t}\nonumber \\
\times & e^{-\frac{q_0^2 + r^2}{2\sigma^2}}
\underbrace{\int_{0}^{2\pi}\cos(m\phi-m\Omega\tau)e^{-i(2n-1)\phi}\,d\phi}_{=\pi\delta_{m,2n-1}\exp(-im\Omega\tau)}\,r^2 dr\nonumber \\
& =
\frac{1}{\sigma^{2}}\sum_{n=1}^{\infty}\int_{0}^{\infty}
I_{2n-1}\left[\frac{q_0 r}{\sigma^2}\right]J_n(2\chi g_0 r^2t) e^{-\frac{q_0^2 + r^2}{2\sigma^{2}}}\nonumber \\
& \times\exp[i\Omega(t-(2n-1)\tau)]\,r^{2}dr,
\end{align}
where the sum was restricted to $n\geq1$ since we are interested
in echoes at $t>\tau$.

For a reasonable approximation of the first echo response at $t=\tau$,
we can set $J_{n}(2\chi g_0 r^2 t)\rightarrow J_{n}(2\chi g_0 q_0^2 t)$,
and use the asymptotic expansion of $I_{2n-1}(x)=e^{-x}/\sqrt{2\pi x}$,
such that
\begin{align}
\braket{q_{+}(t\approx\tau)} 
& \approx
\frac{1}{\sigma\sqrt{2\pi q_0}}\int_0^{\infty}e^{-\frac{(r-q_0)^2}{2\sigma^2}}
J_1(2\chi g_0 q_0^2 t)\nonumber \\
&\times
\cos[\Omega(t-(2n-1)\tau)]\,r^{3/2}dr\nonumber \\
&\approx
\frac{q_{0}}{\sigma\sqrt{2\pi}}J_{1}(2\chi g_0q_0^2t)
\int_0^{\infty}e^{-\frac{(r-q_0)^2}{2\sigma^2}}\nonumber \\
&\times
\cos[\Omega(t-(2n-1)\tau)]dr\nonumber \\
\approx q_0 J_1(2\chi g_0 q_0^2 t) & e^{-2q_0^2\sigma^2\chi^2 t^2}\cos[(1+\chi q_0^2)t],
\end{align}
where we used the result from Eq.~\eqref{eq:App-average-position}.
The approximate result shows that the amplitude of the first echo
is given by $q_0 J_1(2\chi g_0 q_0^2\tau)$.

\section{Free Quantum Dynamics of a Coherent State \label{sec:App-Quantum-Dynamics}}

The freely evolving annihilation operator $\hat{a}(t)$ satisfies $\dot{\hat{a}}=i[\mathcal{\hat{H}},\hat{a}]=-i(\hat{I}+2\chi\hat{n})\hat{a}$, i.e.,
\begin{equation} \label{eq:App-a(t)-free-solution}
\hat{a}(t)=e^{-it}e^{-i2\chi\hat{n}t}\hat{a}.
\end{equation}
To evaluate the expectation value in Eq.~\eqref{eq:quantum-<q>-free},
we need to know the action of $\hat{a}(t)$ on a coherent state,
\begin{align}
\hat{a}(t)\ket{\alpha_0} 
&= e^{-it}e^{-i2\chi\hat{n}t}\hat{a}\ket{\alpha_0} 
= \alpha_0e^{-it}e^{-i2\chi\hat{n}t}\ket{\alpha_0}\nonumber \\
&= \alpha_0 e^{-it}\ket{\alpha_0 e^{-i2\chi t}},
\end{align}
where we used the fact that an exponential of a number operator acting on a coherent state
is a scaling transformation, i.e., 
$\exp(\lambda\hat{n})\ket{\alpha}=\ket{\exp(\lambda)\alpha}$.
Next, we use the identity
\begin{equation} \label{eq:App-coherent-states-overlap}
\braket{\beta|\alpha}=\exp\left[-\frac{1}{2}(|\beta|^2 + |\alpha|^2-2\beta^*\alpha)\right],
\end{equation}
such that
\begin{align}
\braket{\hat{a}(t)} 
& =\alpha_0 e^{-it} \braket{\alpha_0|\alpha_0e^{-i2\chi t}} \nonumber \\
& =\alpha_0 e^{-it} e^{-|\alpha_0|^2}\exp\left(|\alpha_0|^2e^{-i2\chi t}\right).
\end{align}
Combining the above, we arrive at a handy formula
\begin{equation} \label{eq:App-handy-formula}
\braket{\beta|\hat{a}(t)|\alpha}=\beta e^{-it}e^{-(|\alpha|^2+|\beta|^2)/2}
\exp\left[\alpha^{*}\beta e^{-i2\chi t}\right].
\end{equation}

\textbf{\emph{Numerical simulation.}} For numerical simulations, 
Eq.~\eqref{eq:dimensionless-quantum-H} is represented in the Fock basis 
and solved with the Python package \texttt{QuTiP}~\citep{johansson2012qutip}.
Like in the classical numerical simulations, the delta impulse $g(t)$ was approximated
by a Gaussian function, see  Eq.~\eqref{eq:App-Gauss-gt}.
The time-step ($\Delta t$) for solving ODE is set to $\Delta t=10^{-5}$. 

\section{Quantum Dynamics of a Kicked Coherent State \label{sec:App-Quantum-Dynamics-Kicked}}

Here, we derive the closed-form formula for the position expectation value of a kicked 
Kerr oscillator, initialized in a coherent state, $\ket{\alpha_0}$.\\

\textbf{\emph{Before the kick.}}
The time-evolved coherent state driven by the Kerr Hamiltonian is given by
\begin{equation}
\ket{\psi(t)}=e^{-\frac{1}{2}|\alpha_0|^2} \sum_{n=0}^{\infty}\frac{\alpha_0^n}{\sqrt{n!}}
e^{-i\left[\chi n(n-1)+n+\frac{1}{2}\right]t}\ket{n}.
\end{equation}

Appendix~\ref{sec:App-Decomposition-Coherent-States} shows that at 
$t=T_\text{rev}/\nu = 2\pi/(\chi \nu)$ (i.e., at $\nu$-th fractional revival), the initial
coherent state, $\ket{\alpha}$ is an \emph{exact} superposition of $\nu$ (or $\nu/2$ if 
$\nu$ is even) coherent states evenly spaced on a circle of radius $\propto|\alpha_0|$ in
the `phase space'
\begin{equation} \label{eq:App-wf-at-T/nu}
\begin{aligned}\ket{\psi(T/\nu)}= & \sum_{k=0}^{\nu-1}C_k\ket{\alpha_k},\\
\alpha_{k} =& \alpha_0 e^{-2\pi i/(\chi \nu)}e^{2\pi ik/\nu},\\
     C_{k} =& \frac{1}{\nu}\sum_{l=0}^{\nu-1}\exp\left\{ -\frac{2\pi i}{\nu}[l^2 + l(k-1)]\right\} .
\end{aligned}
\end{equation}

\textbf{\emph{Action of the kick.}}
The impulsive kick is modeled by the unitary operator, $\hat{U}=\exp[ig_0(\hat{a} + \hat{a}^{\dagger})^2/2]$. For $g_0\ll1$, $\hat{U}$ can be approximated by a product
of a squeezing operator and a phase rotation. For small $g_0$, only the squeezing part
has a significant effect, thus
\begin{equation}
\hat{U}\approx \exp\left\{ i\frac{g_{0}}{2}[\hat{a}^{2}+(\hat{a}^{\dagger})^{2}]\right\}.
\end{equation}
This operator is equivalent to the squeezing transformation \cite{WallsMilburnBook2007}:
\begin{equation}
\begin{aligned}
\hat{U} & =\hat{S}(-ig_0),\\
\hat{S}(\kappa) & =\exp\left\{ \frac{1}{2}[\kappa^*\hat{a}^2 - \kappa(\hat{a}^{\dagger})^{2}]\right\}.
\end{aligned}
\end{equation}
The action of $\hat{U}$ on a coherent state $\ket{\alpha}$ can be approximated using the
identity $\hat{S}(\kappa)\ket{\alpha}=\hat{S}(\kappa)\hat{D}(\alpha)\ket{0}
=\hat{D}[\cosh(|\kappa|)\alpha - e^{i\theta}\sinh(|\kappa|)\alpha^*]\hat{S}(\kappa)\ket{0}$,
where $\kappa=|\kappa|e^{i\theta}$. For $|\kappa|\ll1$, we can neglect the vacuum squeezing,
such that
\begin{equation}
\begin{aligned}
\hat{U}\ket{\alpha} &\approx \ket{\bar{\alpha}},\quad \text{where} \\
\quad \bar{\alpha}  & = \cosh(g_0)\alpha+i\sinh(g_0)\alpha^* \approx \alpha+ig_0\alpha^*.
\end{aligned}
\end{equation}

\textbf{\emph{Post-Kick Evolution.}}
The kick transforms the wave function in Eq.~\eqref{eq:App-wf-at-T/nu} into
\begin{equation}
\hat{U}\ket{\psi(T/\nu)} = \sum_{k=0}^{\nu-1} C_k \ket{\bar{\alpha}_k}.
\end{equation}
To compute the position expectation value, 
$\langle\hat{q}(t)\rangle=\sqrt{2}\text{Re}\left[\langle\hat{a}(t)\rangle\right]$,
we use Eq.~\eqref{eq:App-handy-formula}, such that (for $t>\tau$)
\begin{align}
\langle\hat{a}(t)\rangle 
&=
\sum_{k,k'=0}^{\nu-1} C_k^* C_{k'}
\langle\bar{\alpha}_k|\hat{a}(t-\tau)|\bar{\alpha}_{k'}\rangle\nonumber \\
=e^{-i(t-\tau)} 
&\sum_{k,k'=0}^{\nu-1}C_k^*C_{k'}\bar{\alpha}_{k'}
\exp\left[-\frac{|\bar{\alpha}_{k}|^{2}+|\bar{\alpha}_{k'}|^{2}}{2}\right]\nonumber \\
&\times\exp\left[\bar{\alpha}_k^*\bar{\alpha}_{k'}e^{-i2\chi(t-\tau)}\right].
\label{eq:App-full-analytic}
\end{align}

\begin{widetext}
The signal can be written as
\begin{equation}
\braket{\hat{a}(t)} = e^{-i(t-\tau)}\sum_{l=0}^{\nu-1}\sum_{k=0}^{\nu-1}
C_k^* C_{k+l}\bar{\alpha}_{k+l}
\exp\left[-\frac{|\bar{\alpha}_k|^2 + |\bar{\alpha}_{k+l}|^2}{2}\right]
\exp\left[\bar{\alpha}_k^*\bar{\alpha}_{k+l}e^{-i2\chi(t-\tau)}\right],
\end{equation}
because all involved functions of $k$ are $\nu$ periodic.
\end{widetext}

\textbf{\emph{Explicit form.}}
The expansion coefficients $C_{k}$ in Eq.~\eqref{eq:App-wf-at-T/nu} are generalized
Gauss sums. Depending on the number-theoretical properties of $\nu$, the sum has a closed-form
solution. To keep things simple, let's choose $\nu$ odd, then
\begin{equation}
C_{k}=\begin{cases}
\frac{1}{\sqrt{\nu}}e^{i2\pi\frac{(\nu+1)}{4\nu}(k-1)^2}, & \nu\,\text{mod}\,=1,\\
-\frac{i}{\sqrt{\nu}}e^{i2\pi\frac{(\nu+1)}{4\nu}(k-1)^2}, & \nu\,\text{mod}\,4=3.
\end{cases}\label{eq:C_k_analytical}
\end{equation}

Next, we simplify
\begin{widetext}
\begin{align}
\frac{|\bar{\alpha}_k|^2 + |\bar{\alpha}_{k+l}|^2}{2} 
&=
\alpha_0^2\left[1 - g_0^2 + g_0
\left(\sin\left[\frac{4\pi(\chi(k+l)-1)}{\nu\chi}\right]
+\sin\left[\frac{4\pi(k\chi-1)}{\nu\chi}\right]\right)\right]\nonumber \\
&=\alpha_{0}^{2}\left[1-g_0^2
+
2g_0 \sin\left[\frac{4\pi k}{\nu}+\frac{2\pi}{\nu}(l-2/\chi)\right]
\cos\left[\frac{2\pi l}{\nu}\right]\right]\nonumber \\
&=\alpha_0^2\left[1+2g_0\sin\left[\tau(\chi l-2)\right]
\cos(l\chi\tau)\right]+\mathcal{O}(g_0^2).
\end{align}
Likewise, the product $\bar{\alpha}_k^* \bar{\alpha}_{k+m}$ reads
\begin{equation}
\bar{\alpha}_k^*\bar{\alpha}_{k+m} = \alpha_0^2
\left[e^{\frac{2i\pi m}{\nu}}
+
2g_0\sin\left[\frac{4\pi k}{\nu}
+
\frac{2\pi}{\nu}(m-2/\chi)\right]\right]
+
\mathcal{O}(g_0^2).
\end{equation}
\end{widetext}
Combining the results so far, and additionally

\begin{itemize}
\item approximating $\bar{\alpha}_{k+l}\approx\alpha_{k+l}$ (i.e., neglecting the small
amplitude correction of the order of $g_0$ of the overall scaling factor of 
$\braket{\hat{a}(t-\tau)}_{l}$) 
\item assuming $\nu \; \text{mod} \; 4=3$ (e.g., $\nu=23$) for concreteness [this choice selects the second formula in Eq.~\eqref{eq:C_k_analytical}]
\end{itemize}
we arrive at
\begin{widetext}
\begin{align}
\braket{\hat{a}(t)}_{l} 
& \approx
\alpha_0 e^{-\alpha_0^2-it}
\exp\left[i\pi l\left(\frac{\nu+1}{2\nu}l-\frac{\nu-1}{\nu}\right)\right]
\exp\left[\alpha_0^2e^{\frac{2i\pi l}{\nu}}e^{-i2\chi(t-\tau)}\right]\nonumber \\
\times\frac{1}{\nu} & \sum_{k=0}^{\nu-1}
\left\{ \exp\left[\frac{\pi il(\nu+1)k}{\nu} + \frac{2\pi ik}{\nu}\right]
\exp\left\{ 2\alpha_0^2 g_0\left[e^{-i2\chi(t-\tau)}
-
\cos\left(\frac{2\pi l}{\nu}\right)\right]\right.\left.\sin\left[\frac{4\pi k}{\nu}+\frac{2\pi}{\nu}\left(l - \frac{2}{\chi}\right)\right]\right\}\right\}.
\label{eq:isolated_pulses_simplification}
\end{align}

The sum appearing in Eq.~\eqref{eq:isolated_pulses_simplification} can be reduced to
a single term as follows. First, we use the Jacobi-Anger expansion
\begin{align}
\exp\left\{ \underbrace{2\alpha_0^2g_0\left[e^{-i2\chi(t-\tau)}-\cos\left(\frac{2\pi l}{\nu}\right)\right]}_{=z(t)}
\sin\left[\frac{4\pi k}{\nu}+\frac{2\pi}{\nu}\left(l-\frac{2}{\chi}\right)\right]\right\} \nonumber \\
=\sum_{r=-\infty}^{\infty}J_r[z(t)/i]
\exp\left[ir\left[\frac{4\pi k}{\nu}+\frac{2\pi}{\nu}\left(l-\frac{2}{\chi}\right)\right]\right].
\end{align}
Then,
\begin{align}
\frac{1}{\nu}\sum_{k=0}^{\nu-1}\exp\left[\frac{\pi il(\nu+1)k}{\nu}+\frac{2\pi ik}{\nu}\right]
\sum_{r=-\infty}^{\infty}J_{r}[z(t)/i]
\exp\left[ir\left[\frac{4\pi k}{\nu}+\frac{2\pi}{\nu}\left(l-\frac{2}{\chi}\right)\right]\right]
\nonumber \\
=\frac{1}{\nu}\sum_{r}J_{r}[z(t)/i]e^{i\frac{2\pi r}{\nu}\left(l-\frac{2}{\chi}\right)}
\sum_{k=0}^{\nu-1}\exp\left[i\frac{2\pi}{\nu}k\left(\frac{l(\nu+1)}{2}+1+2r\right)\right].
\end{align}
\end{widetext}

The sum over $k$ is a delta function and thus defines a `selection rule',
\begin{equation}
\frac{1}{\nu}\sum_k\exp\left[i\frac{2\pi}{\nu}k\left(\frac{l(\nu+1)}{2}+1+2r\right)\right]
=
\delta_{r=r^*},
\end{equation}
and $r^{*}$ is a solution of 
\begin{equation} \label{eq:selection_rule_r}
\frac{l(\nu+1)}{2}+1+2r=0, \quad \text{mod}\;\nu.
\end{equation}
Recall that $l$ is a non-negative integer, $l=0,1,2,\dots,\nu-1$, and $r$ is an integer.
Every choice of $r$ fixes $l$, as shown in Table~\ref{tab:example-selection-rule} for $\nu=23$.
\begin{table}
\begin{centering}
\begin{tabular}{c|c|c}
$r$ & $l$ & $\tau+\tau l/2$\tabularnewline
\hline 
$-3$ & 10 & $6\tau$\tabularnewline
\hline 
$-2$ & 6 & $4\tau$\tabularnewline
\hline 
$-1$ & 2 & $2\tau$\tabularnewline
\hline 
$0$ & 21 & $\nu\tau/2=T_\text{rev}$\tabularnewline
\hline 
$1$ & 17 & $\pi-2\tau$\tabularnewline
\hline 
$2$ & 13 & $\pi-4\tau$\tabularnewline
\hline 
3 & 9 & $\pi-6\tau$\tabularnewline
\end{tabular}
\par\end{centering}
\caption{Example of the selection rule on $l$. Here, $\nu=23$ for concreteness.
\label{tab:example-selection-rule}}
\end{table}
Next, Eq.~\eqref{eq:isolated_pulses_simplification} reduces to
\begin{align}
\langle\hat{a}(t)\rangle_{r_{*}} & \approx\alpha_{0}e^{-\alpha_{0}^{2}}e^{-i(t+2r_{*}\tau)}J_{r_{*}}[z(t)/i]\nonumber \\
 & \times\exp\left[\alpha_{0}^{2}e^{-i2\chi(t-\tau-l\tau/2)}\right],
\end{align}
where we used the fact that 
\begin{equation}
\exp\left[\frac{i2\pi l}{\nu}r_*
+
i\pi l\left(\frac{l(\nu+1)}{2\nu}-\frac{\nu-1}{\nu}\right)\right]=1.
\end{equation}
It follows from Table~\ref{tab:example-selection-rule} that $\tau+\tau l/2$ equals $-2r_*\tau$.

For convenience, we flip the sign of $r_*$ (now, $l$ in $z(t)$ is given by 
$l=4r^*-2\text{ mod }\nu$)
\begin{align}
\braket{\hat{a}(t>\tau)}_l
&\approx
\alpha_0e^{-\alpha_0^2}e^{-i(t-2r_* \tau)}J_{r_*}[iz(t)]\nonumber \\
&\times\exp\left[\alpha_{0}^{2}e^{-i2\chi(t-2r_*\tau)}\right],
\end{align}
such that $r_* = 1,2,3,\dots$ selects the `classical echo' of order $r_*$, while 
$r_* = -1,-2,-3$, selects the `quantum echo' of order $|r_*|$. Note that 
$\braket{\hat{a}(t)}_l/J_{r^*}[iz(t)]$ has the same time dependence as in 
Eq.~\eqref{eq:quantum-<q>-free}.
For small $\tau$ (relative to $T_\text{rev}$), the classical echoes overlap with the 
low-order echoes obtained in the classical system.

 \section{Exact Decomposition at the Fractional Revival \label{sec:App-Decomposition-Coherent-States}}
At $t = T/\nu = 2\pi/(\chi \nu)$, the evolved state of the system is given by
\begin{equation} \label{eq:psi_frac_rev}
\ket{\psi(T/\nu)}=e^{-|\alpha_0'|^2/2}\sum_{n=0}^{\infty}
\frac{(\alpha_0')^n}{\sqrt{n!}}e^{-\frac{2\pi i}{\nu} n(n-1)}\ket{n}.
\end{equation}
The rotated coherent amplitude is defined as
\begin{equation}
\alpha_0'=\alpha_0e^{-2\pi i/(\chi \nu)}
\end{equation}
and it accounts for the evolution under the harmonic oscillator Hamiltonian,
$\hat{n}+\hat{I}/2$.

To express the state in terms of coherent states, we insert a discrete Fourier resolution
of the identity:
\begin{equation}
\delta_{n\equiv l \,\text{mod}\, \nu}=\frac{1}{\nu}\sum_{k=0}^{\nu-1}e^{\frac{2\pi i}{\nu}k(n-l)}.
\end{equation}
Inserting $1=\sum_{l=0}^{\nu-1}\delta_{n\equiv l \; \text{mod}\; \nu}$ into 
Eq.~\eqref{eq:psi_frac_rev}, we get
\begin{align}
\ket{\psi(T/\nu)} & =e^{-|\alpha_0'|^2/2}
\sum_{n=0}^{\infty}\frac{(\alpha_0')^n}{\sqrt{n!}}e^{-\frac{2\pi i}{\nu}n(n-1)}\nonumber \\
&\times
\overbrace{\sum_{l=0}^{\nu-1}\frac{1}{\nu}\sum_{k=0}^{\nu-1}e^{\frac{2\pi i}{\nu}k(n-l)}}
^{=\sum_{l=0}^{\nu-1}\delta_{n\equiv l \,\text{mod}\, \nu}=1}\ket{n}\nonumber \\
& =\frac{e^{-|\alpha_0'|^2/2}}{\nu}\sum_{k=0}^{\nu-1}\sum_{l=0}^{\nu-1}e^{-2\pi i\frac{l(l-1)+kl}{\nu}}\nonumber \\
&\times\underbrace{\sum_{n=0}^{\infty}\frac{(\alpha_0'e^{2\pi ik/\nu})^n}{\sqrt{n!}}\ket{n}}_{\equiv\ket{\alpha_k}}.
\end{align}
The last sum is a coherent state:
\begin{equation}
\ket{\alpha_{k}} = \Ket{\alpha_0'e^{\frac{2\pi i}{\nu}k}}
=e^{-\frac{|\alpha_0'|^2}{2}}\sum_{n=0}^{\infty}\frac{(\alpha_0'e^{\frac{2\pi i}{\nu}k})^n}{\sqrt{n!}}\ket{n}.
\end{equation}
To summarize, at the $\nu$-th fractional revival, the single initial coherent state, 
$\ket{\alpha}$ turns into a superposition $\nu$ coherent states evenly spaced on a circle of
radius $\propto|\alpha_0|$ in the `phase space'
\begin{equation}
\begin{aligned}
\ket{\psi(T/\nu)} &= \sum_{k=0}^{\nu-1}C_{k}\ket{\alpha_k},\\
\alpha_{k}      &= \alpha_{0}'e^{2\pi ik/\nu} = \alpha_0e^{-2\pi i/(\chi \nu)}e^{2\pi ik/\nu} \\
     C_{k}      &= \frac{1}{\nu}\sum_{l=0}^{\nu-1}\exp\left[-\frac{2\pi i}{\nu}[l^2 + l(k-1)]\right].
\end{aligned}
\end{equation}

\section{Free Dynamics of a Cat State \label{sec:App-cat-state-dynamics}}

In this subsection, we derive Eq.~\eqref{eq:cat-<q(t)>-explicit}.
First, we find the normalization constant for a two-component cat state 
$\ket{\mathcal{C}}=\mathcal{N}_+\ket{+\alpha_0} + \mathcal{N}_-e^{i\theta}\ket{-\alpha_0}$
using Eq.~\eqref{eq:App-coherent-states-overlap}
\begin{equation}
\mathcal{D}^{2}\equiv\braket{\mathcal{C}|\mathcal{C}}=\mathcal{N}_{+}^{2}+\mathcal{N}_{-}^{2}+2\mathcal{N}_{+}\mathcal{N}_{-}e^{-2|\alpha_{0}|^{2}}\cos{\theta}.
\end{equation}
Then,
\begin{align}
\braket{\mathcal{C}|\hat{a}|\mathcal{C}}
&=\mathcal{N}_+^2\braket{+\alpha_0|\hat{a}|+\alpha_0}
+\mathcal{N}_-^2\braket{+\alpha_0|\hat{a}|+\alpha_0}\nonumber \\
+\mathcal{N}_+\mathcal{N}_-&(e^{i\theta}\braket{+\alpha_0|\hat{a}|-\alpha_0}
+e^{-i\theta}\braket{-\alpha_0|\hat{a}|+\alpha_0}).
\end{align}
Using Eq.~\eqref{eq:App-handy-formula}, this simplifies to
\begin{align}
\braket{\mathcal{C}|\hat{a}|\mathcal{C}} &= \alpha_0 e^{-it}e^{-|\alpha_0|^2} 
\{(\mathcal{N}_+^2-\mathcal{N}_-^2)\exp\left[|\alpha_0|^{2}e^{-i2\chi t}\right]\nonumber \\
&-i2\mathcal{N}_+\mathcal{N}_-  \sin(\theta)\exp[-|\alpha_0|^2e^{-i2\chi t}]\}.
\end{align}
Finally, $\braket{\hat{q}(t)}=(\sqrt{2}/\mathcal{D}^{2}){\rm Re}[\braket{\mathcal{C}|\hat{a}|\mathcal{C}}]$.
\section{Dissipation \label{sec:App-Dissipation}}

The analytical solution of the master equation in Eq.~\eqref{eq:master-equation}
\textit{without the kick} is given by
\begin{equation}
\hat{\rho}(t)=e^{\hat{\mathcal{L}}t}=e^{\hat{S}t}e^{\hat{L}t}\exp\left[\frac{1-e^{-t(\gamma+i2\chi\hat{R})}}{\gamma+i2\chi\hat{R}}\hat{J}\right]\hat{\rho}(0),
\end{equation}
where $\hat{S}$, $\hat{L}$ and $\hat{J}$ are superoperators defined by
\begin{equation}
\begin{aligned}
\hat{S}\hat{\rho} & =-i\chi[(\hat{a}^\dagger)^2\hat{a}^2,\hat{\rho}],\quad
\hat{L}\hat{\rho}  =-\frac{\gamma}{2}(\hat{a}^{\dagger}\hat{a}\hat{\rho}+\hat{\rho}\hat{a}^{\dagger}\hat{a}),\\
\hat{R}\hat{\rho} & =\hat{a}^\dagger \hat{a}\hat{\rho} - \hat{\rho}\hat{a}^\dagger\hat{a},
\quad
\hat{J}\hat{\rho} =\gamma\hat{a}\hat{\rho}\hat{a}^{\dagger}.
\end{aligned}
\end{equation}
To propagate an operator instead of the density matrix, we have to act with 
$\exp(\hat{\mathcal{L}}^\dagger t)$. Thus, $\hat{a}(t)$ is given by
\begin{equation}
\hat{a}(t)=\exp\left[\frac{1-e^{-t(\gamma-i2\chi\hat{R})}}{\gamma-i2\chi\hat{R}}\hat{J}^\dagger\right]
e^{\hat{L}^\dagger t}e^{\hat{S}^\dagger t}\hat{a},
\end{equation}
where the order of exponential operators has been reversed. Applying $\exp(S^\dagger t)$
first, we get
\begin{equation}
\hat{a}(t) \!=\! e^{-it}\exp\left[\frac{1-e^{-t(\gamma-i2\chi\hat{R})}}{\gamma-i2\chi\hat{R}}\hat{J}^\dagger\right]
e^{\hat{L}^\dagger t}e^{-2i\chi t\hat{n}}\hat{a}.
\end{equation}
Next, we verify the action of $\exp(\hat{L}^\dagger t)$ on a product of any function of 
$\hat{n}$ and $\hat{a}$:
\begin{align}
\hat{L}^\dagger g(\hat{n})\hat{a} 
 & =-\frac{\gamma}{2}[\hat{n}g(\hat{n})\hat{a}
   +g(\hat{n})\hat{a}\hat{n}]\nonumber \\
 & =-\frac{\gamma}{2}[\hat{n}g(\hat{n})\hat{a}+g(\hat{n})(\hat{n}\hat{a}+\hat{a})]\nonumber \\
 & =-\frac{\gamma}{2}[\hat{n}g(\hat{n})\hat{a}+\hat{n}g(\hat{n})\hat{a}+g(\hat{n})\hat{a}]\nonumber \\
 & =-\frac{\gamma}{2}[2\hat{n}+\hat{I}]g(\hat{n})\hat{a}.
\end{align}
Thus,
\begin{equation}
e^{\hat{L}^\dagger t}e^{-2i\chi t\hat{n}}\hat{a}
=
e^{-(\gamma/2)[2\hat{n}+\hat{I}]t}e^{-2i\chi t\hat{n}}\hat{a}.
\end{equation}
Likewise, it can be shown that the operator $\hat{R}$ is equivalent to $-1$ in this case, and 
$\hat{J}^\dagger$ is equivalent to $\hat{n}$.
Finally, the Heisenberg annihilation operator at zero temperature is given by 
\begin{align}
\hat{a}(t) 
=&
e^{-it}e^{-\frac{\gamma t}{2}}
\exp\left[\frac{1-e^{-t(\gamma-i2\chi\hat{R})}}{\gamma-i2\chi\hat{R}}\hat{J}^\dagger\right]
e^{-(2i\chi+\gamma)\hat{n}t}\hat{a}\nonumber \\
= e^{-it} & e^{-\frac{\gamma t}{2}}\exp\left[\frac{1-e^{-(2i\chi+\gamma)t}}
{2i\chi+\gamma}\gamma\hat{n}\right]
e^{-(2i\chi+\gamma)\hat{n}t}\hat{a}.
\end{align}
When $\gamma \ll 1$, the main effect of the damping is captured by the simplified
expression
\begin{equation}
\hat{a}(t) \approx e^{-it} e^{-(2i\chi+\gamma)\hat{n}t}\hat{a}.
\end{equation}
This solution can be obtained from $\hat{a}(t)$ in Eq.~\eqref{eq:App-a(t)-free-solution},
$\hat{a}(t)=e^{-it}e^{-i2\chi\hat{n}t}\hat{a}$, by the substitution 
$2i\chi \rightarrow 2i\chi +\gamma$.

\end{document}